\documentclass[twocolumn,showpacs]{revtex4}
\usepackage{epsfig}
\usepackage{amsmath}

\begin{document}
\title{Study of spin models with polyhedral symmetry on square lattice} 
\author{Tasrief  Surungan$^1$}
\email{tasrief@unhas.ac.id}
\author{Yutaka Okabe$^2$}
\email{okabe@phys.metro-u.ac.jp}
\affiliation{$^1$Department of Physics, Hasanuddin University, Makassar, South Sulawesi 90245, Indonesia\\
$^2$Department of Physics, Tokyo Metropolitan University, Hachioji, 
Tokyo 192-0397, Japan }
\date{19 September 2012}

\begin{abstract}
Anisotropy is important for the existence of true long range order 
 in two dimensional (2D) systems.
This is firmly exemplified by the $q$-state clock models 
in which discreteness drives the quasi-long range order into
a true long range order at low temperature for $q> 4$.
Previously we studied 2D edge-cubic spin model,
which is one of the discrete counterpart of the 
continuous Heisenberg model, and observed two finite
 temperature phase transitions, each corresponds
to the breakdown of octahedral ($O_h$) symmetry and 
$C_{3h}$ symmetry, which finally freezes into ground 
state configuration.
The present study investigates discret models with polyhedral 
 symmetry, obtained by e
equally partioning the $4\pi$ of the solid angle
of a sphere.  There are five types of models if spins are only
allowed to point  to the vertices of the
polyhedral structures such as Tetrahedron, Octahedron, Hexahedron, 
Icosahedron and Dodecahedron.
By using  Monte Carlo simulation with cluster algorithm we 
calculate order parameters and estimate the critical temperatures
exponents of each model. We found a systematic decrease in critical 
temperatures as the number of spin states increases (from
the Tetrahedral to Dodecahedral spin model).
\end{abstract}

\maketitle

\section{Introduction}
Phase transitions are  ubiquitious phenomena in nature, firmly exemplified by
the melting of ice, spontaneous magnetization of
ferromagnetic material and transformation from normal 
conductor of metal into a superconductivity
at very low temperatures. In general, a phase transition is
 related to the breakdown of symmetry of a system\cite{Landau}.
For a thermal-driven phase transition, systems are in high degree 
of symmetry at high temperature because all configurational spaces 
are accessible.  The decrease in temperature will reduce
thermal fluctuation and the system stays in some favorable states.
If the phase transition occurs with no latent heat, the system experiences
continuous transition, also known as second  order phase transition,
which is a transition between the ordered and the disordered state.

According to Mermin-Wagner-Hohenberg theorem, 
spin models with continuous symmetry and short-range interaction
can not have a true long range order (TLRO) for two dimensional (2D) 
lattices, thus no finite temperature transition\cite{Mermin}.
However, a unique type transition called  Kosterlitz-Thouless 
(KT) transition can exist in the XY model (O(2) symmetry)\cite{kosterlitz}.
It is a transition between a high temperature paramagnetic phase
and a low-temperature quasi-long range order (QLRO), known as KT phase.
If the $2\pi$ planar angle 
 of the XY model is discretized into $q$ equal angles, 
we obtain a $q$-state Clock model. This model, apart from inheriting 
the KT phase, possesses a lower-temperature TLRO
driven by the discretness\cite{jose,ts04}.

It is of interest to systematically study the role played by the 
discrete symmetry for  3D case. In analogy with the Clock models for
2D symmetry, we discretize the continuous orientation of Heisenberg 
spin (O(3) symmetry) for obtaining spin models with polyhedral symmetry.
This is done by equally partioning the $4\pi$ solid 
angle of a sphere, resulting in five regular  polyhedrons, 
also known as Platonic solids, i.e., Tetrahedron, Octahedron, Cube, 
Icosahedron and Dodecahedron\cite{MacLean}.  Table \ref{Table01} tabulates the 
characteristics of each structure, to which we define a model with 
spin orientations restricted to point to the vertices of the 
corresponding structure.  Previously we study the edge-cubic spin model with
 underlying symmetry, the Octahedral symmetry ($O_h$), similar to 
that of Hexahedron and Octahedron (cubic) model\cite{ts06}. 
However, spin orientation of the model is only allowed to point to the 
middle point of cubic's edges, therefore there are 12 possible states. 
We observed two finite temperature phase transitions which comes from the fact 
that this model partitions the solid angle unequally.

The present paper studies models with polyhedral symmetry.
We expect to observe finite temperature second order 
phase transitions due to the discreteness.
\begin{table} 
\caption{Characteristics of regular polyhedrons.}\label{Table01}
\begin{center}
\begin{tabular}{lcccc}
\hline
\hline
 Name & Vertices & Faces & Edges & Group Symmetry\\
  & ($q$-state) &  &  & \\
\hline
Tetrahedron   & 4  & 4 &  6 & $S_4$ \\
Octahedron   & 6  & 8 & 12 & $O_h = S_4\times C_2$ \\
Hexahedron (cube)   & 8  & 6 &  12 & $O_h$ \\
icosahedron   & 12  & 20 & 30& $A_5\times C_2$  \\
dodecahedron   & 20  & 12 & 30 & $A_5\times C_2$  \\
\hline
\end{tabular}
\end{center}
\end{table} 
The remaining part of the paper is organized
as follows: Section II describe the model and the method. 
The result  is  discussed in Section III. Section IV is devoted
to the summary and concluding remark.  

\section{Model and Simulation Method}

The polyhedral spin models are the discrete version of the Heisenberg
model with spins are only allowed to point to the vertices of the structures 
listed in Table \ref{Table01}. The Hamiltonian of the model
is written as follows
\begin{equation}
H = - J \sum_{\langle ij \rangle}\vec s_i \cdot \vec s_j 
\end{equation}
where $\vec s_i$ is the spin on site $i$-th. Summation is performed 
over all the nearest-neighbor pairs of spins on a square lattice with
ferromagnetic interaction ($J > 0$) and with periodic boundary condition. 
The energy of the ground state configuration, i.e., when all spins having a common
orientation, is $-2NJ$  with $N$ is the number of spins.

\begin{figure}[t]
\includegraphics[width=1.0\linewidth]{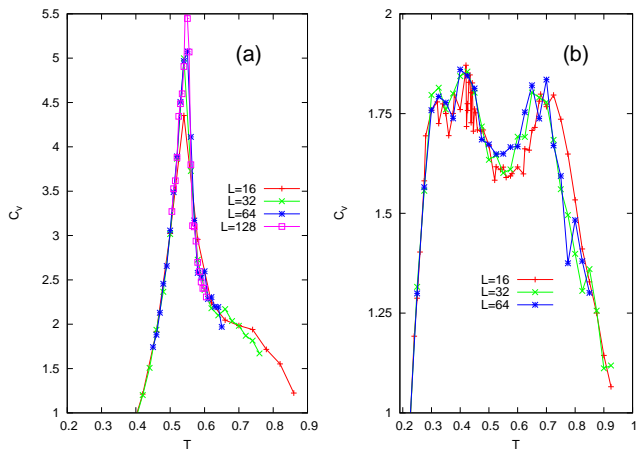}
\vspace{0.1cm}
\caption{The temperature dependence of the specific heat for
various system sizes of (a) Dodecahedron and Icosahedron models.
 As shown, there exists a clear peak for Icosahedron
while two peaks for dodecahedron.
The peaks may signify the existence of phase transitions.
 The error bar, in average, is in the order of
symbol size.}
\label{SPHT}
\vspace{0.6cm}
\end{figure}

We use the canonical Monte Carlo (MC) method with  single
 cluster spin updates introduced by Wolff \cite{wolff} and 
adopt Wolff's idea of embedded scheme in constructing a cluster
for the 3D vector spins. Spins are projected
into a randomly generated plane so that they 
are divided into two Ising-like spin groups.  
This scheme is essential for carrying out cluster algorithm applied to
such spins as 2D and 3D continuous  spins.

After the projection, the usual steps of the cluster algorithm is 
 performed \cite{kasteleyn},
i.e., by connecting bonds from the randomly
chosen spin to its nearest neighbors of similar group, 
with suitable probability. 
This procedure is repeated for neighboring spins
connected to a chosen spin until no more spins to include.
One Monte Carlo step (MCS) is defined as visiting once the whole 
spins randomly and perform cluster spin update in each visit.
It is to be noticed that for each step
 a spin may be updated many times, in average,
 in particular near the critical point.

Measurement is performed after enough equilibration MCS's ($10^4$ MCS's). 
Each data point is obtained from the average over several parallel
runs, each run is of $4 \times 10^4$ MCS's. 
To evaluate the statistical error each run is treated as a single measurement.
For the accuracy in the estimate of critical exponents and 
temperatures,  data are collected  upto
more than 100 measurements for each system size.

\section{Results and Discussion}

\subsection{Specific heat and magnetization}

The first step in the search  for any possible phase transition
is to measure the specific heat defined as follows

\begin{equation}
C_v(T) = \frac{1}{Nk_BT^2}(\langle E^2\rangle - \langle E\rangle^2)
\end{equation}
where $E$ is the energy in unit of $J$  
 while $\langle \cdots \rangle$ represents
the ensemble average of the corresponding quantity.
All temperatures are expressed in unit of $J/k_B$ where
$k_B$ is the Boltzmann constant.

\begin{figure}
\includegraphics[width=1.0\linewidth]{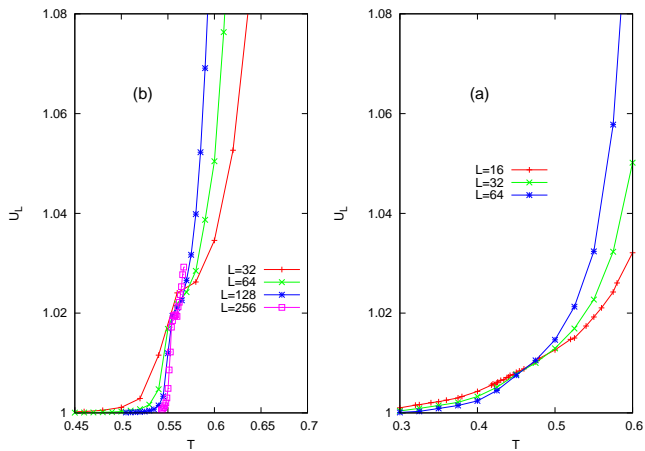}
\vspace{0.1cm}
\caption{Temperature dependence of moment ratios
 for several system sizes of (a) Icosahedron and (b)
Dodecahedron models. 
The crossing points indicates a phase transition between
the disordered  and the intermediate phase. 
Error bar is in the order smaller than the symbol size.}
\label{RATIO_m}
\vspace{0.5cm}
\end{figure}

The specific heats of Dodecahedron and Icosahedron models
are ploted in Fig. \ref{SPHT}. 
Although peaks in a specific heat are more directly related 
to energy fluctuation,
they may signify the existence of  phase transitions.
More quantitative analysis in searching for phase transition is performed 
through the evaluation of the order parameters from which
critical temperatures and exponents may be extracted using  finite size 
scaling (FSS) procedure.
In this paper we present the
analysis of obtaining exponents only for 
 Dodecahedron and Icosahedron models as other models are 
equivalent to the commonly known models.
The Tetrahedron model is equivalent to the 4-state Potts model while
the Hexahedron (corner-cubic model) is equivalent to the Ising model.
The Octahedron model which is face-cubic model has been 
studied by Yasuda and Okabe\cite{yasuda}.

As the probed system is ferromagnetic we consider magnetization
$M = |\sum \vec s_i|$ as the order parameter. By defining 
$M^k$ as the $k$-th order moment of magnetization and
$g(R) = \sum \vec s(r) \cdot \vec s(r+R)$ as correlation function,
the moment and correlation ratios are respectively written as 
follows
\begin{eqnarray} 
U_L &=&\frac{\langle M^4\rangle}{\langle M^2\rangle^2}\\
Q_L &=&\frac{\langle g(L/2)\rangle}{\langle g(L/4)\rangle}\\
\nonumber
\end{eqnarray}  
Precisely, the distance $R$ for the correlation function $g(R)$
is a vector quantity. Here we take the simple and  more
convenient distances, i.e.,  $L/2$ and $L/4$, both in $x$- and $y$-directions.

\begin{figure}
\includegraphics[width=1.0\linewidth]{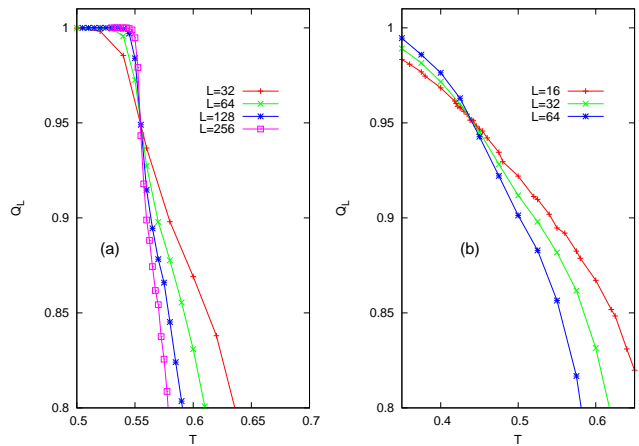}
\vspace{0.1cm}
\caption{Temperature dependence of correlation ratios
 for several system sizes of (a) Icosahedron and (b)
Dodecahedron models.  The absis of the crossing points 
are the critical temperatures of the corresponding models, 
comparable to the numerical values given by the moment ratio
of Fig. \ref{RATIO_m}.  Error bar is in the order smaller than 
the symbol size.}
\label{RATIO_c}
\vspace{0.5cm}
\end{figure}

\begin{figure}
\includegraphics[width=1.0\linewidth]{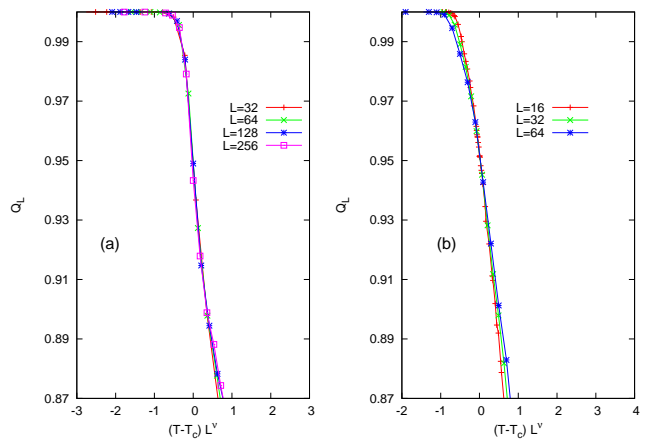}
\vspace{0.1cm}
\caption{The FSS plot of correlation ratio 
for (a) Icosahedorn and (b) Dodecahedron models. 
The estimates of critical temperature and the exponent of correlation
length $\nu$ are obtained.}
\label{FSS_c}
\vspace{0.5cm}
\end{figure}

The existence  of a phase transition can be observed from the 
temperature dependence
of $U_L$ and $Q_L$. At very low temperature where system is approaching 
the ground state, both moment and correlation ratio are trivial.
 Due to the absence of fluctuation,
 the distribution of $M$ is a delta-like function,
  giving moment ratio equals to unity.
Correlation ratio also goes to unity as correlation
function for small and large distance is the same 
due to  highly correlated state. 
In excited states, the moment and the correlation
ratios are not trivial, they depend on temperature.
The plot of moment ratio for various system sizes of
Icosahedron and Dodecahedron models shown
in Fig. \ref{RATIO_m}, exhibits  crossing points indicating
phase transitions. The crossing point for
the Icosahedron model is slightly mild compared to
the that of Dodecahedron which is related to the
performance of moment ratio. Crossing
points for both models are strongly indicated
by the plot of correlation ratio shown in Fig. \ref{RATIO_c}.
The procedure for estimating  critical temperatures
using FSS will be presented in the next subsection.

\subsection{Finite Size Scaling}
FSS analysis for obtaining critical temperature and 
exponents are shown in Fig. \ref{FSS_c}, where we  plot  correlation ratio 
of the models.  In general, moment ratio has larger correction
to scaling than the correlation ratio \cite{tomita02a,tasrief05}, which happens
to be the case here, shown for example by the mild crossing point 
of moment ratio for Icosahedron models (Fig. 2a), while
sharp crossing for correlation ratio (Fig. 2b).
However, if the variables of the two correlation functions
are not local  quantity,  the correlation ratio may have 
larger correction to scaling.
Our estimate of $T_c$ is based on result obtained
from the correlation ratio. 
 For Icosahedron model, the estimated values of $T_c$ and $\nu$ are 
respectively 0.555(1) and 1.30(1), while for Dedecahedron, $T_c=0.438(1)$
and $\nu = 2.01(1)$.
The number in bracket is the uncertainty in the last digit.

\begin{figure}
\begin{center}
\includegraphics[width=0.9\linewidth,height=1.1\linewidth]{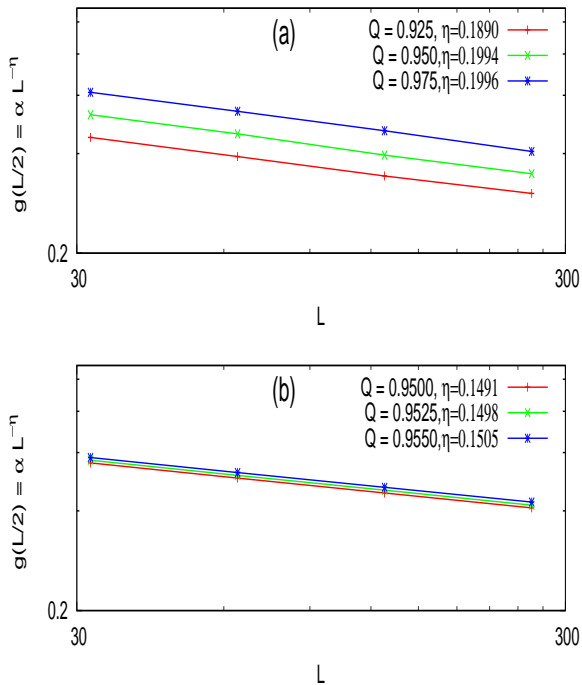}
\vspace{0.5cm}
\caption{Double logarithmic plot of $\bar g(L/2)$ vs $L$. The
gradient of the fitted line associated with $Q=0.82$
for Icosahedron and $Q=0.92$ for Dodecahedron model. The best estimate
for $\eta$ are respectively $0.149(1)$ and $0.199(1)$.}
\label{Eta}
\vspace{0.4cm}
\end{center}
\end{figure}

Using the correlation ratio we can also extract
the decay exponent $\eta$ of the correlation function.
This is done by  looking at the constant value of correlation ratio $Q$
for different sizes and then find the corresponding correlation function $g(L/2)$.
The correlation function is in power-law dependence on 
the system size, $g(L/2) \sim L^{-\eta}$ \cite{tomita02a}.
Therefore, if we plot $g(L/2)$ versus $L$ for various $Q$'s
in logarithmic scale, as in Fig.~\ref{Eta},
the value of $\eta$ will correspond to the gradient of
the best-fitted line for each constant of correlation ratio.
There are several lines plotted in Fig. \ref{Eta}. Since
the critical temperature is associated with the value of $Q \sim 0.95$
for Icosahedron model
(Fig. \ref{RATIO_c}(a)), we assign $\eta = 0.199(1)$ as the best estimate. 
For the Dodecahedron model (Fig. \ref{Eta}b) the estimate in $\eta=0.149(1)$.

After obtaining the critical exponents, we can now discuss 
the universality classes of the existing phase transitions.
The expectation that models with the same underlying
symmetry has to belong to the same universality class
seems to be too good to apply. As indicated, although
the underlying symmetry of the Icosahedorn and the Dodecahedron
is the same, both models  have different universality
class. It is of interest to investigata whether
this finding also holds for 3D systems.

\section{Summary and Concluding Remarks}\label{four}

In summary, we have studied 
critical properties of spin models with polyhedral symmetry
on a square lattice. They are the discrete
version of the Heisenberg model.
If the $4\pi$ solid angle is equally
partitioned, then there exist five 
regular octahedrons, as listed in Table \ref{Table02}.
We only consider the Icosahedron and the Dodecahedron
models as the Tetrahdron and the octahedron
are equivalent to  the common
models, i.e., the Ising and the 4-state Potts
model, respectively while  the Hexahedron  model has
been studied by Yasuda and Okabe.
We observed a second order phase transition 
for each correspoding model studied
and  estimated the critical temperature and exponents
by using FSS of correlation ratio. 
Our results are tabulated in Table \ref{Table02}, including 
results from previous studies. 
We found  a systematic decrease in critical 
temperatures as the number of spin states 
increases ($q \to \infty$ as $ T_c \to 0$). This implies that
$T_c =0$ for the model with continuous symmetry, which
emphasizes the importance of discretness in 2D  systems.

\begin{table}
\caption{Critical temperatures and exponents
 of phase transitions of 2D Polyhedral spin models.$\\$}
\label{Table02}
\begin{tabular}{c|c|c|c}
\hline
\hline
Model & $T_c$ & $\nu$& $\eta$\\
($q$-state) &  & & \\
\hline
4 & $(4/3)*(1/\ln(3))=0.214$& 2/3 & 1/4\\
6 &$0.9085(2)$& 0.685(2) &  0.23(1)\\
8 &$(1/3)*(2/\ln(2.42))=0.756$& 1 &  1/4\\
12 & $T_c = 0.555(2)$& $\nu = 1.31(1)$ & 0.199(1) \\
20& $T_c = 0.438(1)$& $\nu = 2.0(1)$ & 0.149(1) \\
\hline
12$^*$\cite{ts08}& $0.602(1)$& 1.50(1) &  0.260(1)\\
\hline
\end{tabular}
\end{table}

\section*{Acknowledgments}

The authors wish to thank J. Kusuma and Bansawang BJ for valuable discussions.  
 The extensive computation was performed using the  supercomputer facilities of
 the Institute of Solid State Physics, University of Tokyo, Japan and
Parallel computers at the Department of Physics, Hasanuddin 
University.
The present work is financially supported by the Incentive Research Grant
No. 246/M/Kp/XI/2010 of Indonesian Ministry of Research and Technology.


\begin{thebibliography}{99}
\bibitem{Landau} L. D. Landau, {\it On the theory of phase transition}, in {\it Collected Paper of L. D. Landau}, edited by D. T. Haar (Pergamon Press, 1965).
\bibitem{Mermin}N. D. Mermin and H. Wagner, Phys. Rev. Lett. {\bf 17}
1133, (1966) ; P. C. Hohenberg, Phys. Rev. {\bf 158}, 383 (1967).
\bibitem{kosterlitz} J. M. Kosterlitz and D. Thouless, J. Phys. C {\bf 6},
1181 (1973); J. M. Kosterlitz, J. Phys. C {\bf 7}, 1046 (1974).
\bibitem{MacLean} K. J. M. MacLean, {\it 
A Geometric Analysis of the Platonic Solids and Other Semi-Regular Polyhedra}
,  (The Big Pictures. Press, Oxford, 2002).
\bibitem{jose} J. V. Jos\'e, L. P. Kadanoff, S. Kirkpatrick, and D. R. Nelson,
 Phys. Rev. B {\bf 16}, 1217 (1977).
\bibitem{ts04} T. Surungan, Y. Okabe, and Y. Tomita,  J. Phys. A {\bf 37}, 4219 (2004).
\bibitem{ts08}Tasrief Surungan, Naoki Kawashima and  Yutaka Okabe, Phys. Rev. B77, 214401 (2008).
\bibitem{aharony}A. Aharony, Phys. Rev. B {\bf 10}, 3006 (1974).
\bibitem{kim}D. Kim, P. M. Levy and L. F. Uffer, Phys. Rev. B {\bf 12}, 989 (1975).
\bibitem{sznajd}J. Sznajd and M. Dudzi\'nski,  Phys. Rev. B {\bf 59}, 4176 (1999).
\bibitem{calabrese02} P. Calabrese and A. Celi,  Phys. Rev. B {\bf 66}, 184410 (2002).
\bibitem{carmona} J. M. Carmona, A. Pelissetto, and E. Vicari,  Phys. Rev. B {\bf 61}, 15136 (2000).
\bibitem{calabrese04} P. Calabrese, E. V. Orlov, D. V. Pakhnin and A. I. Sokolov, Phys. Rev. B {\bf 70}, 094425 (2004).
\bibitem{yasuda}T. Yasuda and Y. Okabe, {\it in preparation}.
\bibitem{ashkin} J. Ashkin and E. Teller, Phys. Rev. {\bf 64}, 178 (1943).
\bibitem{wolff} U. Wolff,  Phys. Rev. Lett. {\bf 62}, 361 (1989).
\bibitem{kasteleyn} P. W. Kasteleyn and C. M. Fortuin, J. Phys. Soc. Jpn, Suppl. {\bf 26}, 11 (1969);
 C. M. Fortuin and P. W. Kasteleyn, Physica (Amsterdam) {\bf 57}, 536, (1972). 
\bibitem{tomita02a} Y. Tomita and  Y. Okabe,  Phys. Rev. B {\bf 66}, 180401(R) (2002).
\bibitem{tasrief05} T. Surungan and Y. Okabe,  Phys. Rev. B{\bf 71}, 188428 (2005).
\bibitem{wu} F. Y. Wu, Rev. Mod. Phys. {\bf 54}, 235, (1982).

\end{thebibliography}
\end{document}